\documentclass[twocolumn,showpacs,preprintnumbers,pre,fleqn]{revtex4}
\usepackage{epsfig}
\usepackage{graphicx}
\usepackage{amsfonts}
\newcommand{\ig}{\includegraphics}
\newcommand{\ct}{\cite}

\newcommand{\D}{\Delta}

\newcommand{\beas}{\begin{eqnarray*}}
\newcommand{\eeas}{\end{eqnarray*}}
\begin{document}
                                                                                

\title{Effect of discontinuity in threshold distribution on the critical behaviour of a random fiber bundle}
\author{Uma Divakaran}
\email{udiva@iitk.ac.in}
\author{Amit Dutta}
\email{dutta@iitk.ac.in}
\affiliation{Department of Physics, Indian Institute of Technology Kanpur - 208016, India}
\date{\today}
\begin{abstract}

The critical behaviour of a Random Fiber Bundle Model with mixed uniform 
distribution of threshold strengths and global load sharing rule is studied 
with a special emphasis on the nature of distribution of  
avalanches for different parameters of the distribution. 
The  discontinuity in the threshold strength distribution of fibers  
non-trivially 
modifies the critical stress as well as puts a restriction on the allowed 
values of parameters for which the recursive dynamics approach holds good.
The discontinuity 
leads to a non-universal behaviour in the avalanche size
distribution for smaller values of avalanche size.
We observe that apart from the mean field behaviour for larger avalanches,
a new behaviour for smaller avalanche size is observed 
as a critical threshold distribution
 is approached. 
The phenomenological
understanding of the above result is provided using the 
exact analytical result for the avalanche size distribution. Most interestingly,
the prominence of non-universal behaviour in avalanche size distribution 
depends on the system parameters.  
\end{abstract}
\maketitle
\section{Introduction}
Breakdown phenomena in nature has captured the attention of scientists
for years\cite{benguigui}. A study of this phenomena plays a major role
for the prediction of failure and design of materials and structures.
One of the paradigmatic model mimicking the fracture processes
is random fiber bundle model (RFBM) which is simple yet subtle enough
to capture the essential physics of the breakdown phenomena. 

Random fiber bundle models
 \cite{silveria,reviewchak,dynamic,moreno} have been studied
extensively in recent years. Typically a RFBM consists of $N$ 
parallel fibers with randomly
distributed threshold strength $(\sigma_{th})$ taken from a given distribution. 
If the stress generated due to an external force is greater than $\sigma_{th}$
of a fiber, it breaks.
The dynamics of the model is initiated by applying a small external force 
just enough to break the weakest fiber present in the bundle. The load carried
by this broken fiber is shared amongst the remaining intact fibers 
following a load sharing rule causing further failures.
When no further failure takes place, the external force is once again 
increased quasistatically to break the weakest intact 
element present in the 
bundle and the process continues till
the bundle breaks down completely at an external stress called
 the critical stress. Even though the threshold distribution of
real materials may not be known exactly, in theoretical models
the distributions are usually approximated by either a uniform
distribution  or a Weibull distribution \ct{reviewchak}.

The avalanche size is defined as the number of broken fibers between two
successive loadings.
The distribution of avalanche size turns out to be a key factor
in charaterising any breakdown  phenomena. 
Hemmer and Hansen\cite{hemmer}
studied the avalanche size distribution $D(\Delta)$ of an avalanche of
size $\Delta$ in a RFBM with the global load sharing (GLS) scheme, in which 
the additional stress due to a broken fiber 
is distributed equally to the remaining 
intact fibers. They  established a universal power-law distribution in the large
$\Delta$ limit given as
$D(\Delta)\propto \Delta^{-\xi}$ with $\xi =5/2$.
\begin{figure}[h]
\includegraphics[height=2.0in]{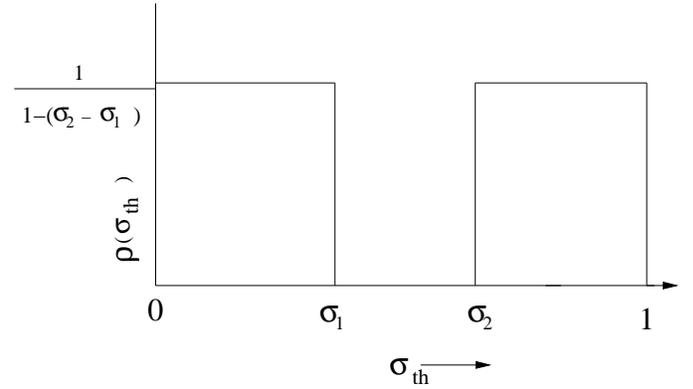}
\caption{Mixed Uniform Distribution}
\end{figure}

RFBM with threshold strengths which are continuous
and are uniformly distributed between 0 to $\sigma_1$ and also between
 $\sigma_2$ to 1
\ct{reviewchak,dynamic},
have been studied separately. But what happens when two such bundles are 
merged is not known, especially  when $\sigma_1<\sigma_2$ such that 
there exists a discontinuity in the threshold strength distribution.
In this paper, we investigate the role of such a discontinuity
on the critical behaviour of a RFBM. 
The distribution of threshold strength of fibers used in the present work
is given as (See Fig.~1)
\begin{eqnarray}
\rho(\sigma_{th}) &=& \frac {1}{1-(\sigma_{2}-\sigma_{1})}~~~~~~
0<\sigma_{th}\leq \sigma_{1}\nonumber\nonumber\\
&=&0~~~~~~~~~~~~~~~~~~~~~~\sigma_{1}<\sigma_{th}<\sigma_{2}\nonumber\\
~~~~~~~~~~&=&\frac {1}{1-(\sigma_{2}-\sigma_{1})}~~~~~\sigma_{2}\leq\sigma_{th}\leq 1.
\end{eqnarray}
The discontinuity as defined above
introduces dilution in the model in the sense that two types of fibers
separated by a gap in their threshold distributions coexist in the same bundle. 
It is shown below that this discontinuity plays a 
crucial role in the dynamics of the model. Here,  
a fraction $f$ of fibers belong to the weaker threshold 
distribution with strengths  uniformly lying between $0$ and $\sigma_1$ 
(Class A) whereas the remaining fraction 
of fibers have stronger threshold strengths between $\sigma_2$ and 1 (Class B).
Clearly, $(\sigma_2 - \sigma_1)$ is the measure of the discontinuity which
vanishes in the limit of purely uniform distribution\ct{dynamic}.

In a recent paper, Pradhan, Hansen and Hemmer  \cite{prl} showed that 
for a 
bundle which is close to the complete break down ($i.e.,$ imminent failure), 
a crossover in $\xi$ from a value  5/2 to 3/2 is observed 
when the threshold distribution approaches the critical distribution.
Critical distribution in their case is the distribution in which 
the lowest threshold of 
the remaining intact fibers is equal to half that of the strongest.
This crossover has also been observed with  other load sharing rules\ct{kun06}.
We are however interested in looking at the distribution 
of total avalanche size 
and a crossover from a power-law behaviour with $\xi=5/2$ to
a non-mean field, non-universal behaviour for smaller $\Delta$ 
is observed near a critical distribution. 
This interesting behaviour is due to the presence of class A fibers.
In the present model, the critical distribution corresponds to 
$\sigma_2=0.5$. We emphasize that in Ref.~[7], the proximity of $x_0$ 
(the threshold of the weakest intact fiber)
to the the critical threshold (=0.5) leads to the crossover behaviour
whereas in the present case it is the proximity of $\sigma_2$ to the
critical value (0.5) which is at the root of the observed crossover behaviour.
In a sense, $\sigma_2$ is playing a role analogous to $x_0$ in our model.

The paper is organized as follows: We have already introduced 
the model above.
The critical stress and exponents are obtained in Sect. II(A) with a special 
emphasis on the avalanche size distribution in Sect. II(B). The concluding 
remarks are presented in Sect. III.

\section{Results and Discussions}
\subsection{Critical stress and exponents}

To study the dynamics of failure of fibers, we use the recursive dynamics 
approach \cite{dynamic}.
If a fraction $f$ of the total fibers belong to Class A and the remaining 
$1-f$ to class B, then the uniformity of the distribution demands
\begin{eqnarray}
f =\frac{1}{1-(\sigma_{2}-\sigma_{1})}\int_{0}^{\sigma_{1}} d\sigma ;\nonumber\\
~~~ {\rm so~~ that},~~\sigma_{1} = \frac{f}{1-f}(1-\sigma_{2}).
\end{eqnarray}

The above equation provides a relationship between $f$, $\sigma_1$ and 
$\sigma_2$ and at the same time puts a restriction on the allowed values
of the parameter $\sigma_1$ as shown below. 
Any value of  $\sigma_1>f$
leads to  a value of $\sigma_2$ smaller than $\sigma_1$ 
which is not an acceptable
distribution (see Eq.~1). 

We now define $U_t$ as the 
fraction of unbroken 
fibers after a time step $t$. Then the redistributed stress at the instant $t$
is $\sigma_t=F/N_t=\sigma/U_t$
where the applied force $F=N\sigma$ and $N_t=NU_t$. 
The recurrence relations between $U_t,U_{t+1}$ and between 
$\sigma_t,\sigma_{t+1}$ for the GLS are obtained as \cite {dynamic}:
\begin{eqnarray}
U_{t+1} &=& 1-P(\sigma_{t})=1-P(\frac{\sigma}{U_{t}})\nonumber\\
{\rm and}~\sigma_{t+1} &=& \frac{\sigma}{U_{t+1}} =
\frac {\sigma} {(1-P(\sigma_t))}
\end{eqnarray}
\noindent where $P(\sigma_t)$ is the fraction of broken fibers with the
redistributed stress $\sigma_t$, and is given as
$$P(\sigma_t)=\int_0^{\sigma_t}\rho(\sigma_{th})d\sigma_{th}.$$
This dynamics propagates until no further breaking takes place. 
It should be emphasised that the initial load is so small that 
the redistributed stress is always less than $\sigma_2$ and therefore
 fibers from class B  cannot  fail. 
Thus to initiate the breaking of class B fibers,
the redistributed stress at a later time $t$ must 
exceed $\sigma_2$. The fixed point solution for $U$ ($=U^*$) and 
$\sigma(=\sigma^*)$ at which no further failure takes place
can be obtained using the standard technique of solving the above recursive
relations. By substituting $P(\sigma_t)$ in Eq.~(3) we get
\begin{eqnarray}
U_{t+1} &=& 1-P(\frac{\sigma}{U_{t}}) \nonumber \\
 &=& 1-[\frac{\sigma_{1}}{1-(\sigma_{2}-\sigma_{1})} +\frac{1}{1-(\sigma_{2}-\sigma_{1})} (\frac{\sigma}{U_{t}}-\sigma_{2})]\nonumber.
\end{eqnarray}
At the fixed point,
$$ U^{*}= 1-[\frac{\sigma_{1}}{1-(\sigma_{2}-\sigma_{1})} +\frac{1}{1-(\sigma_{2}-\sigma_{1})} (\frac{\sigma}{U^{*}}-\sigma_{2})] $$
giving the following stable fixed point solutions
\begin{eqnarray}
U^{*} &=& \frac {1}{2(1-(\sigma_{2}-\sigma_{1}))} [1 + \sqrt
{1-\frac{\sigma}{\sigma_{c}}]}~~{\rm and}\nonumber\\
\sigma^{*} &=& \frac 1 {2} - \frac 1 {2} \sqrt { 1 - \frac {\sigma}{\sigma_c}}
 ~~~~~~~~{\rm with}\\ 
\sigma_{c} &=& \frac {1} {4[1-(\sigma_{2}-\sigma_{1})]}. 
\end{eqnarray}
If the external 
applied stress is less than $\sigma_c$, the system reaches a fixed point.
For $\sigma > \sigma_c$, the
bundle breaks down completely as both the $U^*$ and $\sigma^*$ 
become imaginary.
Equation (4) suggests that  the redistributed stress $\sigma^*$ attains the
maximum value (=0.5) at $\sigma_c$. 
The critical stress of the mixed model  varies with the gap
($\sigma_2 - \sigma_1$) and reduces to the value $\sigma_c = 1/4$ for the
uniform distribution as ($\sigma_2 - \sigma_1) \to 0$. In deriving Eq.~(4), 
$\sigma_t$ is assumed to be greater than $\sigma_2$ 
while the applied stress when plotted against the redistributed 
stress shows a discontinuity at $\sigma_1$. Beyond $\sigma_1$,
 the external force 
is increased to break the fiber with threshold strength $\sigma_2$,
 i.e., the gap in the threshold distribution
is also reflected in the constitutive behaviour of the model. 

The discontinuity on the other hand imposes some restrictions on the 
parameters for this calculation to be 
valid. Since the maximum value of the redistributed stress is equal to 0.5, 
$\sigma_2$ must be less
than 0.5 
so that some fibers from
class B also fail at the critical point. The condition $\sigma_2 < 0.5$,
eventually
restricts the value of
critical stress to be less than 0.5 for any chosen distribution. 
However, $\sigma_2=0.5$ is a limiting case
when the redistributed stress at the critical point marginally reaches 
class B fibers. In short, we must have
$$ \sigma_1<f, \sigma_2<0.5~.$$
One can define an order parameter $O$ associated with the transition
\cite{dynamic} as shown below:
                                                                                
\begin{eqnarray}
O=2[1-(\sigma_{2}-\sigma_{1})]U^{*} - 1 = (\sigma_{c}-\sigma)^{1/2}=
(\sigma_{c}-\sigma)^{\beta}.\nonumber
\end{eqnarray}
The order parameter goes to 0 as $\sigma \rightarrow \sigma_{c}$
following a power law  $(\sigma_{c}-\sigma)^{1/2}$ .
Susceptibility can be defined as the increment in the number of broken fibers
for an infinitesimal increase of load. Therefore,
\begin{eqnarray}
\chi=\frac{dm}{d\sigma}~~~ {\rm where}~~~ m = N[1-U^{*}(\sigma)].\nonumber
\end{eqnarray}
Hence,
\begin{eqnarray}
\chi \propto (\sigma_{c}-\sigma)^{-1/2}=(\sigma_{c}-\sigma)^{-\gamma}.
\nonumber
\end{eqnarray}
The exponents $\beta$ and $\gamma$ stick to their mean field values
\cite{dynamic,moreno}
$i.e.\beta=\gamma=1/2$ .
We conclude that though the discontinuity alters the critical
stress, the critical exponents remain unaltered.
It is to be noted that the model reduces to the already obtained
results in various limits. For example, if class A fibers are absent 
($f$=0), $\sigma_c=1/4(1-\sigma_2)$ and an elastic to plastic 
deformation is observed \cite {dynamic}.
\subsection{Avalanche size distribution}
Let us now  focus on the avalanche size distribution exponent $\xi$.
Below is shown some of the allowed distributions which 
satisfy the above mentioned restrictions. 
\begin{center}
\begin{tabular}{|c|c|c|c|c|}\hline
Case &~~~$f$~~~&~~~ $\sigma_{1}$~~~~&~~~~$\sigma_{2}$ ~~~~ & ~~~~$\sigma_{c}$\\\hline
1 &0.10  &0.08  &0.28  &0.31\\\hline
2 &0.20  &0.19  &0.24  &0.26\\\hline
3 &0.20  &0.15  &0.40  &0.33\\\hline
4 &0.30  &0.25  &0.42  &0.30\\\hline
5 &0.30&0.29&0.33&0.26\\\hline
6 &0.40&0.35&0.47&0.29\\\hline
\end{tabular}
\end{center}

We study the avalanche size distribution numerically by using
the method of breaking of the weakest fiber\cite{hemmer}. In the simulation,
the fibers are arranged in an increasing order of their threshold strengths.
An external force sufficient to break the weakest fiber is applied
and the load due to the breaking of this fiber gets redistributed among the
remaining intact fibers following the GLS scheme.
The number of failed fibers for a fixed external load is recorded till
the dynamics reaches a fixed point. Thereafter, the external load
is increased further and the above process is repeated till 
the critical stress is reached.

Following interesting observations (Fig.~2)  are clearly highlighted.
(i) For the cases 1, 2 and 5 of the table, the avalanche size exponent
is 5/2. (ii) For the cases 3,4 and 6, there is an apparent power law
behaviour for smaller $\Delta$ with the exponent which
is found to depend on the system parameters. In the examples chosen
here the exponent happens to be close to 3.
For larger $\Delta$, however, we retrieve the universal mean field
behaviour with $\xi=5/2$. Also, (iii) an increase in the
region with $\xi\approx 3$  is seen as $\sigma_2\rightarrow 0.5$. These
observations establish the following: (i) there is a non-universal behaviour
of $D(\Delta)$ in the small $\Delta$ limit, (ii) there is a crossover 
to the universal behaviour in large $\Delta$ limit, and 
(iii) the crossover behaviour is prominent 
as $\sigma_2 \to 0.5$. 

The above mentioned results can be explained by extending the analytical result 
for the avalanche size distribution obtained by Hemmer and Hansen \cite{hemmer}
to the mixed model. The general expression for the avalanche size distribution
with GLS is given as
\begin{eqnarray}
\frac{D(\Delta)}{N}=\frac{\Delta^{\Delta-1}}{\Delta!}\int_0^{x_c}dx\rho(x)
(1-\frac{x\rho(x)}{Q(x)})\nonumber\\
\left[\frac{x\rho(x)}{Q(x)}\right]^{\Delta-1}
{\rm exp}(-\Delta\frac{x\rho(x)}{Q(x)})
\end{eqnarray}
where $x$ is the redistributed stress and $Q(x)$ is the fraction of 
unbroken fibers at $x$. The upper limit of the integration ($x_c$)
is the redistributed stress at the critical point.
The right hand side of Eq.~(6) is broken into two parts, $D_1(\Delta)$ and
$D_2(\Delta)$ for the 
mixed model with any allowed values of $\sigma_1$ and $\sigma_2$ (where
$\rho(x)=\rho=1/(1-\sigma_2+\sigma_1)$):

\noindent As long as the redistributed stress is restricted to the 
class A fibers,
$$Q(x)=\frac{1-\sigma_2+\sigma_1-x}{1-\sigma_2+\sigma_1},$$
we have
\begin{eqnarray}
{D_1(\Delta)}=\frac{\Delta^{\Delta-1}}{\Delta!}\frac{1}
{1-\sigma_2+\sigma_1}\int_0^{\sigma_1}dx
(\frac{1-\sigma_2+\sigma_1-2x}{x})\nonumber\\
\left[\frac{x}{1-\sigma_2+\sigma_1-x}\times
{\rm exp}(-\frac{x}{1-\sigma_2+\sigma_1-x})\right]^{\Delta}.
\end{eqnarray}

\noindent When the redistributed stress belongs 
to the second block (class B), we have
$$Q(x)=\frac{1-x}{1-\sigma_2+\sigma_1}, ~~~{\rm and}$$
\begin{eqnarray}
{D_2(\Delta)}=\frac{\Delta^{\Delta-1}}{\Delta!}
\frac{1}{1-\sigma_2+\sigma_1}\int_{\sigma_2}^
{0.5}dx(\frac{1-2x}{x})\nonumber\\
\times\left[\frac{x}{1-x}
{\rm exp}(-\frac{x}{1-x})\right]^{\Delta}.
\end{eqnarray}

\begin{figure}
\ig[height=3.6in,width=3.6in]{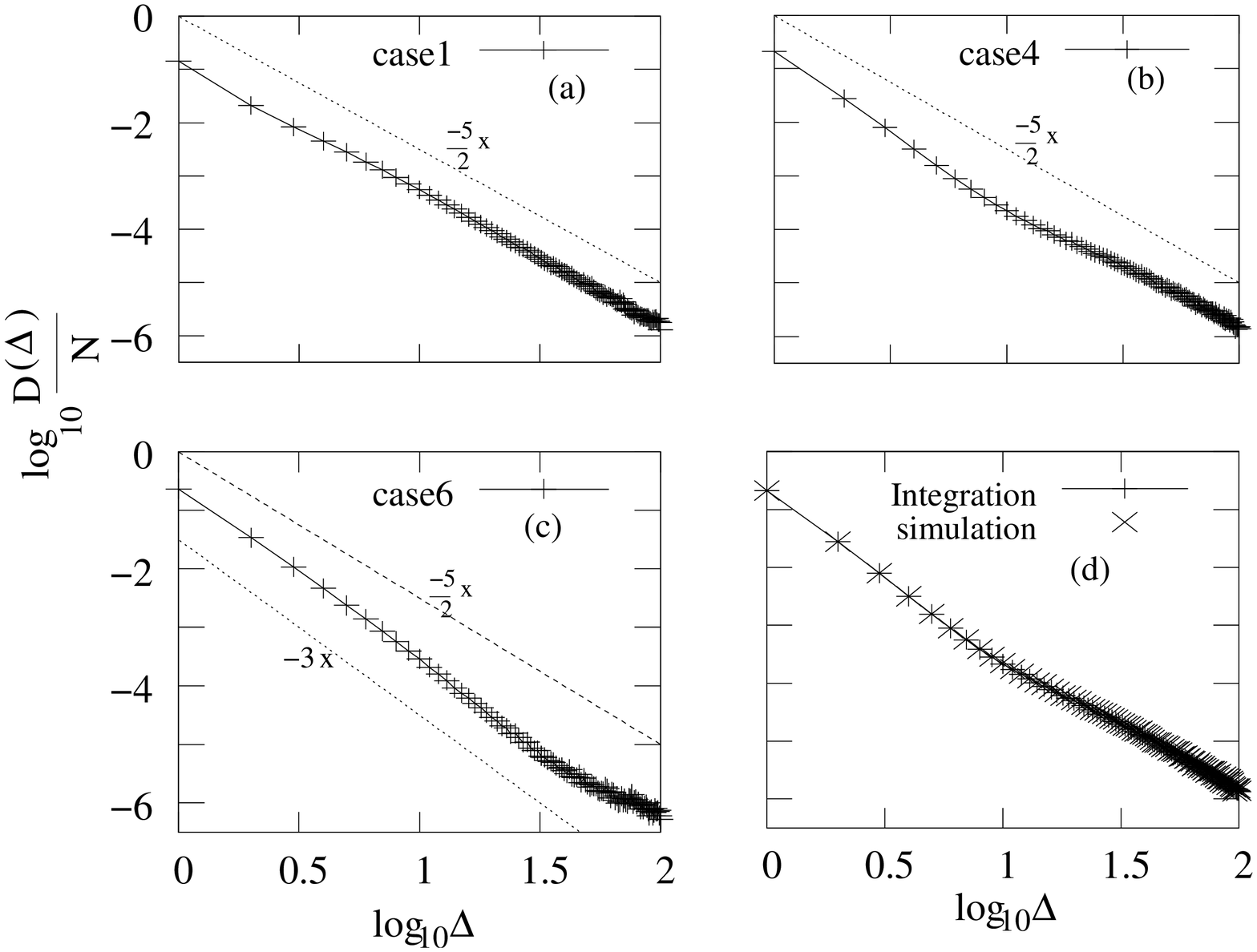}
\caption{
Fig.~(2a) corresponds to case1 of the table 
where no crossover is observed. (2b) shows a crossover from
a non-universal behaviour to the universal behaviour 
with $\xi=5/2$ as $\Delta$ is increased (case4). Fig.~(2c) shows an
increase in region with non-universal behaviour (case6). 
A line with slope -5/2 is drawn in each figure for comparison.
In Fig.~(2c),  a line
with slope -3 is also shown to indicate the apparent power-law behaviour.
Fig.~(2d) compares the simulation and numerical integration data for case4 which
match identically with each other.}
\end{figure}
A close inspection shows that Eq.~(8) corresponds to a situation
where class A fibers are absent and the results match
exactly with Pradhan $et.al$ \cite{prl}
where a crossover in the avalanche
size exponent from 5/2 to a value 3/2 is observed as $\sigma_2\to 0.5$.
Here, the presence of class A 
fibers necessitates the simultaneous study of Eqs.~(7) and (8)
for the total avalanche size distribution. 
For distributions which are closer to the critical distribution, Eq.~(8) yields
exponent $\xi=3/2$ for smaller $\Delta$, 
but the contribution from the Eq.~(7) modifies this small $\Delta$ behaviour.

The results obtained by numerically integrating Eqs.~(7) and (8) match
identically with the numerical simulation results [Fig.~(2d)]. 
We observe a sharp fall of $D_1(\Delta)$ [Eq.~7] 
which points to the fact that 
the avalanches of smaller sizes are contributed mainly by the breaking of 
class A fibers (Fig.~3).
The point of intersection of the two integrals yields 
$\Delta_c$ which is the value of the avalanche size where the  
crossover takes place. The variation of $\Delta_c$ 
for three different cases are shown in Fig.~3. Incidentally in the examples
presented here, $\Delta_c$ increases with $\sigma_2$.
If we look at a particular critical distribution (Fig.~4),
a crossover from an apparent power-law behaviour of $D(\Delta)$ 
with exponent $\xi$ close to 3 for small $\Delta$ 
to a universal behaviour with $\xi=$3/2  is observed. 
\begin{figure}
\ig[height=3.6in,width=3.6in]{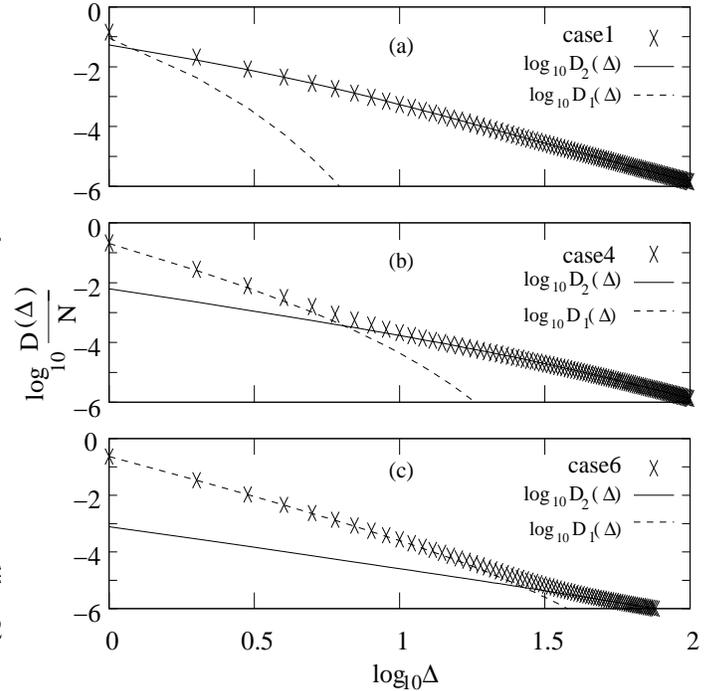}
\caption{Figs.~a, b and c present the numerical integration results showing
an increase in $\Delta_c$ (defined in the text) as $\sigma_2 \to 0.5$. 
The symbol corresponds
to $\log_{10}$D($\Delta$)/N=$\log_{10}(D_1(\D)+D_2(\D))$ whereas 
 $\log_{10}D_1(\D)$ and  
$\log_{10}D_2(\D)$ are drawn to compare their relative magnitudes.}
\end{figure}

We now discuss the numerical integration and simulation results in the light
of Eqs.~(7) and (8). With a change of variable 
$x\rightarrow x/(1-\sigma_2+\sigma_1-x),$ we can rewrite Eq.~(7) 
in the following form
\begin{eqnarray}
{D_1(\Delta)}=\frac{\Delta^{\Delta-1}}{\Delta!}
\int_0^{x_{m}}\frac{1-x}{x(1+x)^2}
e^{(\log x-x)\Delta} dx
\end{eqnarray}
where $x_{m}=\sigma_1/(1-\sigma_2)$.
The argument of the exponential term has a maximum at $x=1$ which is 
outside the range
of integration and hence the saddle point integration method can not be applied
\ct{aval}.
Expanding the term $1/(1+x)^{2}$ in a power series and then using the
incomplete gamma function\cite{abram}, we arrive at the following result 
(see the Appendix)
\begin{eqnarray} 
{D_1(\Delta)}=\frac{e^{(1-x_m)\Delta}}{\Delta^{3/2}}x_m^{\Delta}
\sum_{q=0}^{\infty} (-1)^q(q+1)x_m^q\times \nonumber\\
(\sum_{k=0}^{\infty}
\frac{(x_m\Delta)^k}{(\Delta+q)(\Delta+q+1)...(\Delta+q+k)}\nonumber\\
-\sum_{k=0}^{\infty}
\frac{x_m(x_m\Delta)^k}{(\Delta+q+1)(\Delta+q+2)...(\Delta+q+1+k)}). 
\end{eqnarray}
The leading behaviour of the infinite series (10) is 
$\Delta^{-5/2}e^{(1-x_m)\Delta}x_m^{\Delta}$ 
which justifies the non-universality observed
in numerical simulations in the small $\D$ limit.
The question therefore remains why does the non-universal behaviour
become prominent as $\sigma_2 \to 0.5$. The behaviour of $D_2 (\Delta)$
[Eq.~(8)] for smaller $\Delta (<<\Delta_c)$ is at the root of this. 
When $\sigma_2 \to 0.5$, $D_2 (\Delta)$ goes
as $\Delta^{-3/2}$ and hence the contribution from $D_1(\Delta)$ wins
over to produce a prominent non-universal behaviour. 
The small $\Delta$ behaviour of $D(\Delta)$ is therefore non-mean field 
and non-universal when $\sigma_2 \to 0.5$. On the other hand if $\sigma_2 
<< 0.5$, $x_m$ is relatively smaller and the contribution from class A
fibers decays very fast as $\Delta$ increases and one observes a mean-field
universal behaviour almost for the entire range of $\Delta$. 
We also observe a non-powerlaw behaviour for small $\Delta$ when $\sigma_1$
is very small and $\sigma_2$ is close to criticality as is expected
from the analytical result.
However for a given $\sigma_2$ ($\sim 0.5$), larger $x_m$ leads to a
non-universal behaviour up to larger $\Delta$.
The crossover value $\Delta_c$ is thus roughly given by the value of $\Delta$
for which $D_2(\Delta)$ crosses over from a $\Delta^{-3/2}$ behaviour to
$\Delta^{-5/2}$ behaviour.      
\begin{figure}
\ig[height=2.6in,width=2.6in]{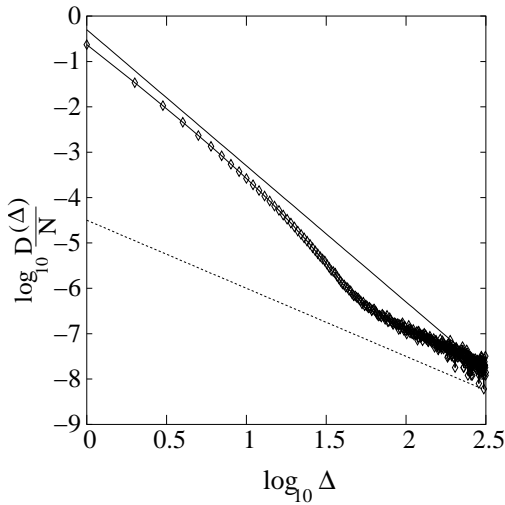}
\caption{Avalanche size distribution for a particular critical threshold
distribution with $f=0.4$. Thick line corresponds to a slope of -3 and dotted
line to -3/2. The asymptotic behaviour clearly is $D(\Delta)\sim\Delta^{-3/2}$.}
\end{figure}

\section{Conclusions}
In conclusion, we have studied a mixed fiber bundle with a discontinuous
but uniform threshold distribution and GLS. Discontinuity leads to a 
functional dependence of
the critical stress on the system parameters $\sigma_1, \sigma_2$ and $f$
and also imposes restrictions on the allowed values of these parameters.
Although the critical exponents are unchanged, there is a non-trivial
change in the burst avalanche distribution behaviour 
where discontinuity leads to a non-universal, non-mean field 
behaviour for small $\Delta$. We would like to emphasise that non-universality
becomes prominent only when $\sigma_2\to 0.5$. For large $\Delta$ limit, the
behaviour is however universal and mean field. If $f=0$ or $\sigma_2<<0.5$, the
non-universal behaviour completely disappears. The non-universality in 
$D(\Delta)$ is also seen for other distributions\cite{aval}. The beauty of
our model is that the non-universal behaviour is tunable with the system
parameters and there is a crossover from non-universality to universality
in the limit of large $\Delta$. One should also note that the imminent failure
of the bundle ($i.e.,$ final stages of the breakdown process) is the same 
as in Ref.\ct{prl} because the effect of class A fibers essentially vanishes
in that limit.

\begin{center}
{\bf Acknowledgments}
\end{center}
The authors thank S. Banerjee, P. Bhattacharyya, B. K. Chakrabarti, Y. Moreno 
S. Pradhan and S. K. Rai for useful help and suggestions. 
\begin{center}
{\bf Appendix}
\end{center}
In this appendix, we shall indicate how to arrive at Eq.~(10) of
the text starting from Eq.~(9).
Equation (9) can be written as
\beas 
D_1(\D) = \frac{e^{\D}}{\D^{3/2}}~ \int_{0}^{x_m}\frac{1-x}{(1+x)^2}~ x^{\D-1}
~e^{-\D x} dx    
\eeas 
\beas 
{\rm Let}~f(\D) &=& \int_{0}^{x_m}\frac{1-x}{(1+x)^2}~ x^{\D-1}
~e^{-\D x} dx    \\
&=&\int_{0}^{x_m} (1-x)~ x^{\D-1}~e^{-\D x}(\sum_{q=0}^{\infty}(-1)^q (q+1)
x^q) dx \\
&=&\sum_{q=0}^{\infty}(-1)^q (q+1)\int_{0}^{x_m} (1-x) x^{\D +q -1}~e^{-\D x} dx. 
\eeas 
The integral can be further written as
\beas 
&& \int_{0}^{x_m} x^{\D +q -1}~e^{-\D x} dx - \int_{0}^{x_m} x^{\D +q}~e^{-\D
x} dx \\
&=& \frac{1}{\D^{\D + q}} \int_{0}^{x_m\D} y^{\D +q -1}~e^{-y} dy \\
& -& \frac{1}{\D^{\D + q+1}}  \int_{0}^{x_m\D} y^{\D +q}~e^{-y} dy. \\
\eeas
Using the power series expansion of Incomplete Gamma Function \cite{abram}
\beas
&=& \frac{1}{\D^{\D + q}} ~~[ e^{-x_m\D} \sum_{k=0}^{\infty} 
\frac{(x_m\D)^{\D+q+k}}{(\D+q)(\D+q+1)\ldots(\D+q+k)}] \\
&&-\frac{1}{\D^{\D + q+1}} \times\\
~~&[&e^{-x_m\D} \sum_{k=0}^{\infty} 
\frac{(x_m\D)^{\D+q+1+k}}{(\D+q+1)(\D+q+2)\ldots(\D+q+1+k)}] \\
\eeas 

Rearranging terms, we get
$$f(\D)=e^{-x_m\D} x_{m}^{\D}  \sum_{q=0}^{\infty} (-1)^q (q+1) x_{m}^q \times $$
$$( \sum_{k=0}^{\infty} \frac{(x_m\D)^k}{(\D+q)(\D+q+1)\ldots(\D+q+k)}$$ 
$$-\sum_{k=0}^{\infty}\frac{x_m~(x_m\D)^k}{(\D+q+1)(\D+q+2)\ldots(\D+q+1+k)}).$$
$i.e.,$ Eq. (10).


\begin{thebibliography}{05}

\bibitem{benguigui}
B. K. Chakrabarti and L. G. Benguigui, {\it Statistical Physics of fracture and
Breakdown in Disordered Systems}, (Oxford University Press, Oxford, 1997);
M. Sahimi, {\it Heterogeneous Materials II: Nonlinear Breakdown Properties and 
Atomistic Modelling}, (Springer-Verlag Heidelberg, 2003);
H. J. Herrmann and S. Roux, {\it Statistical Models of Disordered Media}, (NorthHolland, Amsterdam 1990);

\bibitem{silveria}
F. T. Peirce, J. Text. Inst. {\bf 17}, 355 (1926);
H. E. Daniels, Proc. R. Soc. Londoni, Ser.A {\bf 183} 404 (1945);
B. D. Coleman, J. Appl. Phys. {\bf 29}, 968 (1958);
R. L. Smith, Proc. R. Soc. London, Ser.A {\bf 372}, 539 (1980);
R. da Silveria, Am. J. Phys. {\bf 67}, 1177 (1999);
S. Zapperi, P. Ray, H. E. Stanley, A. Vespignani, Phys. Rev. Lett. {\bf 78}, 1408 (1997);
J. V. Andersen, D. Sornette and K. T. Leung, Phys. Rev. Lett {\bf 78}, 2140
(1997).
\bibitem{reviewchak}
S. Pradhan and B. K. Chakrabarti, Int. J. Mod. Phys. B {\bf 17}, 5565 (2003);
P. C. Hemmer, A. Hansen and S. Pradhan, e-print cond-mat/0602371; in {\it 
Modelling Critical and Catastrophic Phenomena in Geoscience}, edited by
P. Bhattacharya and B. K. Chakrabarti (Springer, Berlin, 2006). p. 27.


\bibitem{dynamic}
S. Pradhan, P. Bhattacharyya and B.K. Chakrabarti, Phys. Rev. E {\bf 66}, 016116
(2002);
P. Bhattacharyya, S. Pradhan and B.K. Chakrabarti, Phys. Rev. E {\bf 67},
046122 (2003).

\bibitem{moreno}
Y. Moreno, J. B. Gomez and A. F. Pacheco, Phys. Rev. Lett. {\bf 85}, 2865 (2000).

\bibitem{hemmer}
P. C. Hemmer and Alex Hansen, J. Appl. Mech. {\bf 59}, 909 (1992);
A. Hansen and P. C. Hemmer, Trends in Stat. Phys. {\bf 1}, 213 (1994);
A. Hansen and P. C. Hemmer, Phys. Lett. A {\bf 184}, 394 (1994).

\bibitem{prl}
S. Pradhan, A. Hansen and P. C. Hemmer, Phys. Rev. Lett. {\bf 95}, 125501 
(2005); S. Pradhan, A. Hansen and P. C. Hemmer, Phys. Rev. E {\bf 74} 016122
(2006);
S. Pradhan and A. Hansen, Phys. Rev. E. {\bf 72}, 026111 (2005).

\bibitem{kun06} 
F. Raischel, F. Kun and H. J. Herrmann, Phys. Rev. E {\bf 74}, 035104 (2006).

\bibitem{aval}
M. Kloster, A. Hansen and P. C. Hemmer, Phys. Rev. E {\bf 56}, 2615 (1997).

\bibitem{abram}
M. Abramowitz and I. A. Stegun, {\it Handbook of Mathematical functions with
Formulas, Graphs and Mathematical Tables} (Courier Dover, New York, 1965).





\end{thebibliography}
\end{document}